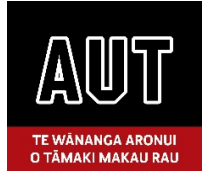

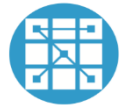

**AUT SOFTWARE ENGINEERING RESEARCH LABORATORY**

# Requirements engineering in global scaled agile software development environment: a multi-vocal literature review protocol

Protocol – Cover Sheet

Technical Report No. SERL_TR_2018_01

**Review Title:** Protocol for a multi-vocal literature review of requirements engineering in global scaled agile software development environment.

**Protocol Developed by:** Priyanka Antil

**Contact Information:**

Priyanka Antil

Software Engineering Research Laboratory [SERL]

School of Engineering, Computer and Mathematical Sciences

Auckland University of Technology

Auckland, New Zealand

priyanka.antil@aut.ac.nz

# Abstract

**Context:** Requirements engineering in global scaled agile software development and the planning phase for a multi-vocal literature review.

**Objective:** Develop a protocol to specify the plan which will be followed to conduct a multi-vocal literature review study on requirements engineering in global scaled agile software development.

**Method:** Kitchenham and Charters (2007), and Garousi et al. (2019) guidelines were followed to develop a protocol for multi-vocal literature review.

**Result:** A validated protocol to conduct a multi-vocal literature review.

**Conclusion:** The review protocol consist of five phases enumerated as follows: research questions, search strategies, validation of review process, reporting the review, and making changes to the protocol.

# 1. Introduction

Scaled agile approaches are observed to be increasing in popularity in the field (Dikert, Paasivaara, & Lassenius, 2016), therefore we conducted an initial brief search of the literature to learn more about requirements engineering (RE) in global scaled agile software development environment. But we could not find a substantial body of academic research on RE in global scaled agile software development environment. Therefore, we decided to conduct a multi-vocal literature review (MLR) as a preliminary study to understand the requirements engineering process, from both practitioner and researcher perspectives in global scaled agile software development environment.

# 2. Research Methodology

An MLR refers to assessment of all possible literature that includes formally published academic literature (e.g., journals, conference papers) as well as unpublished and practitioner literature (e.g., white papers, blogposts) to identify, analyse and interpret the phenomena of interest (Garousi, Felderer, & Mäntylä, 2016)**.** MLR studies can give substantial benefits in certain areas of software engineering (SE) in which new developments are occurring and there is a shortage of academic research (Garousi et al., 2016). Garousi et al. (2016) suggested that an MLR must be conducted as a preliminary study particularly in software engineering because software engineering practitioners produce grey literature (GL) to a large extent, as the most common way of sharing knowledge, advice and experiences on new techniques, approaches and technology driven developments. Otherwise, the researcher could miss out important current information on rapidly evolving real world phenomena of interest.

For this purpose, as suggested by Kitchenham and Charters (2007), we have developed an MLR protocol to specify the plan which will be followed to conduct an MLR study on RE in global scaled agile software development. The key activities covered by the review protocol are described in the following sections (section 3 to section 7).

# 3. Research Questions

The main aim of this MLR is to identify useful information about the software product RE process in a global scaled agile software development environment.  This knowledge in turn will then enable a focused study to be conducted on an area which has had limited attention to date.  This is motivated by an interest in RE, as impacted by the relatively novel phenomenon of adoption of agile methods in large scale global settings. The duration of this MLR study we limit to papers from 2001 as the year in which the agile manifesto was established (Beck et al., 2001).

For that reason, the goals of this MLR are: (i) Identify the challenges of RE in a global scaled agile software development environment, and (ii) Identify effective strategies for surmounting challenges in RE activities in a global scaled agile software development environment.

There are two research questions (RQs) that are set in order to achieve the above goals:

RQ 1: What are the challenges for requirements engineering in global scaled agile software development?

RQ 2: What are the strategies for surmounting challenges in requirements engineering activities in global scaled agile software development?

## 4. Search strategy

The main aim of this step is to define the search and evaluation strategies for classifying the primary studies. The search and evaluation strategies help to perform an exhaustive search of the white literature (WL) and grey literature (GL) that answer the proposed RQs.

### 4.1 Identify search terms

Search terms are identified using the PICO (Population, Intervention, Comparison and Outcomes) framework as recommended by Kitchenham and Charters (2007). The following details of the population, intervention and outcomes will form the basis for the construction of suitable search terms.

**Population:** large software development organizations having globally distributed teams

**Intervention:** scaling agile methods

**Outcome:** understanding of the RE challenges and strategies for surmounting the RE challenges in global scaled agile software development

*Comparison* is not wholly applicable in this work, but is replaced by the notion of *experimental design* covering the range of different approaches in the studies reviewed.

**Experimental design:** empirical studies/evaluative studies, experience reports, blogposts

Section 4.4 (Inclusion criteria) relates to experimental design and our quality assessment criteria (as shown in Section 4.6) covers experimental design in more detail. All primary studies of our MLR study will categorise the experimental design as reported in our spreadsheet metadata under 'Study Type', see section 4.7.

**Major terms:** requirements engineering, global software development, scaled agile.

The search string for this MLR is made up of three substrings: S1, S2, and S3, defined as follows:

S1 is a string which consists of keywords related to requirements engineering practices such as "requirements engineering", "requirements development", "requirements prioritization", "user story", "features", "portfolio management", "backlog management"

S2 is a string consisting of keywords related to global software development/engineering such as "global software development", "Distributed software development"

S3 is a string that consists of keywords related to scale agile software development methods such as "scaled agile framework", "Large-Scale Scrum", "Disciplined Agile Delivery".

Eq. (1). Boolean expression search string

S1 AND S2 AND S3                                                        (1)[1]

In this MLR study, we conducted some initial searches to test and fine-tune the search string. Following is the example of search conducted in the electronic databases: (* - for truncation)

( "requirements engineering" OR "requirements development" OR "requirements prioritization" OR "user story" OR "features" OR "portfolio management" OR "backlog management" ) AND ( "distributed software development" OR global software* ) AND ( "large-scale scrum" OR "scaling agile" OR "disciplined agile delivery" OR "scrum-of-scrum" )

---

[1] We have performed the trial search as well (shown in Appendix A) to train the search string (shown in section 4.1).

## 4.2 Resources to be searched for WL

As per the Kitchenham & Charters (2007) guidelines, three search strategies will be followed to retrieve relevant WL: (i) Automatic search (Digital databases) (ii) Manual Search (Key conferences) (iii) Snowballing

***Automatic Search (Digital databases)***

- IEEE Digital Library,
- ACM Digital Library
- Scopus
- ScienceDirect
- Springer Link

The main reason to include these digital databases was the possibility of accessing their contents (services offered by our institution). Moreover, these databases provide the highest impact full-text journals, conference proceedings, and comprehensively cover the software engineering field in general (Kitchenham & Charters, 2007).

***Manual search***

- Profes - the International Conference on Product-Focused Software Process Improvement: https://www.profes-conferences.org/
- RE - reported as "the premier international scientific venue in the requirements engineering field": https://requirements-engineering.org/
- XP - reported as "the premier Agile software development conference combining research and practice": https://www.agilealliance.org/xp2020/
- ICGSE - The IEEE/ACM International Conference on Global Software Engineering (ICGSE) brings together researchers and practitioners to share their research findings, experiences, and new ideas on diverse themes related to global software engineering.

The reason to include these conferences was because of their standing within the field and specific focus on the key areas of requirements engineering, scaled agile, and global software development.

***Snowballing***

According to the SLR guidelines which are provided by Kitchenham and Charters (2007), forward and backward snowballing will be performed on the selected papers to ensure including all relevant sources as much as possible.

## 4.3 Resources to be searched for Grey Literature

Based on the MLR guidelines which are proposed by Garousi, Felderer, and Mäntylä (2019) the following two different search strategies will be adopted for GL: (i) Automatic search- Google search engine (https://www.google.com), ProQuest Dissertations and Thesis Global database, (ii) Manual search- methods' creator website.

(i) Google search engine: As recommended by Garousi, Felderer, and Mäntylä (2019), an advanced search will be undertaken on regular Google search engine to retrieve relevant master thesis/Ph.D. thesis.

(ii) Digital database: As suggested by Garousi, Felderer, and Mäntylä (2019), an advanced search will be performed on the 'ProQuest Dissertations and Thesis Global' database to retrieve relevant Ph.D. /master thesis.

(iii) Methods' creator websites: According to MLR guidelines which are proposed by Garousi, Felderer, and Mäntylä (2019), a review must include practitioners' website publishing documents (e.g. web pages, reports) on the relevant research area. To retrieve the practitioners' website publishing documents on scaling agile methods, an informal pre-search as suggested by Garousi, Felderer, and Mäntylä (2019) was conducted on a regular Google search engine. Moreover, a set of key consulting firms and scaling agile frameworks who have accompanying support resources on their websites (e.g. case studies, training materials and guides) have been also been identified through $12^{th}$ state of agile report that was produced by VersionOne (2017) and 'a review of scaling agile methods in large software development' that was conducted by Alqudah and Razali (2016). Therefore, the following methods' creator websites will be hand-searched to retrieve potentially relevant documents:

https://www.scaledagileframework.com/

https://less.works/

https://www.disciplinedagiledelivery.com/

https://www.agilealliance.org/resources/experience-reports/

https://www.scrum.org/resources/scaling-scrum

Moreover, (www.stackoverflow.com) *"Stack Overflow is the largest, most trusted online community for developers to learn, share their programming knowledge, and build their careers"* that could be searched for GL in SE (Garousi et al., 2019) but during our informal pre-search, we found most of the information retrieved from "www.stackoverflow.com" was related to programming, or to agile methods practiced at the team level only, and did not extend to large scale use of agile methods. Therefore, this (www.stackoverflow.com) practitioner website will not be included in this MLR study.

### 4.4 Study inclusion and exclusion criteria

The study inclusion and exclusion criteria establish the requirements that a source retrieved from searches must fulfil in order to be included in the study. As suggested by Kitchenham and Charters (2007), the inclusion and exclusion criteria are defined on the basis of the research question(s) as shown in Table 4.1.

| Inclusion criteria | <ul><li>The study is relevant to the search terms as defined in section 3.1.2.</li><li>The study is written in the English language.</li><li>The study is published between 2001 and 2018</li><li>Academic peer reviewed (e.g., journal paper), peer reviewed experience reports (e.g., conference paper), Ph.D. /Master thesis, key consulting firms and scaling agile frameworks who have accompanying support resources on their websites (e.g. case studies, blogposts)</li></ul> |
|---|---|
| Exclusion criteria | <ul><li>Studies that do not focus explicitly on large-scale agile.</li><li>Studies that do not discuss RE in the large-scale agile global setting.</li><li>Studies whose full-text cannot be accessed.</li><li>Duplicate studies (same studies are retrieved from other publications).</li><li>Duplicate studies (same studies in a sequence where only the most relevant or strongest study is included).</li><li>Systematic literature reviews (SLRs) or tertiary studies as these studies would reflect duplicate findings in our primary studies.</li></ul> |

| | • Short papers, panel discussions, PowerPoint presentations, posters and rejected manuscripts. |
|---|---|

Table 4.1: Study inclusion and exclusion criteria

### 4.5 Study selection process

We have defined a separate study selection process for WL and GL by following the Kitchenham and Charters (2007), and Garousi et al. (2019) guidelines.

### 4.5.1 Study selection process for White Literature

**WL_Step1:** In the first step, the search string that was developed in section 4.1 will be applied to all selected digital databases. In the manual search, each set of selected conference proceedings will be examined separately.

**WL_Step 2:** After retrieving the studies from automatic and manual searches, the titles and abstracts of the studies will be examined comprehensively by applying the study inclusion and exclusion criteria (as shown in Table 4.1) to discard the irrelevant studies. If there will be any doubt whether a study should be included or not, that will be discussed with the secondary and third researchers (cf. acknowledgment section below) in order to make an optimal decision.

**WL_Step 3:** In WL_Step 3, the pre-selected studies will be examined based on the full text by applying the study inclusion and exclusion criteria (as shown in Table 4.1).

**WL_Step 4:** Based on the result of automatic and manual searches, snowballing will be performed in WL_Step 4. During this step, WL_Step 2 through 3 will be repeated until no more relevant studies will be found that met the inclusion criteria.

**WL_Step 5:** Primary studies that will be retrieved as a result of WL_Step 3 and WL_Step 4 will be combined, and duplicates will be removed.

### 4.5.2 Study selection process for GL

WL_Step 1 through 3 of screening the WL studies will be repeated in the screening of GL studies.

In GL_Step 4, results will be combined, and the duplicates will be removed to retrieve the final GL studies.

### 4.5.3 Combining final WL studies and GL studies

After performing the WL and GL studies selection process, selected WL studies and GL studies will be combined and the duplicates will be removed.

### 4.6 Study quality assessment criteria

The quality criteria that will be used for assessing the quality of primary studies is adopted from Garousi et al. (2019). These quality criteria are presented in Table 4.2 that cover thoroughness, trustworthiness, and significance of the studies. Each category of quality assessment, which is presented in Table 4.2, will be evaluated on a scale from 0 to 1 except the Publication/literature type. The Publication/literature type category will be evaluated on a scale from 0 to 4 because this is the category where the precedence of the literature is defined.

Note: If the selected study(s) provides a valuable insight, even though the quality score of that study(s) is low, the study(s) will be included in the final pool of the selected of studies.

| Criteria | Questions |
| --- | --- |
| 1. Authority of the publisher (Measure= 0 or 1) | - Is the publishing organization reputable/ Author is associated with reputable organization?<br>- Is the organization/ author cited often by others?<br>- Has the author published other work in the field? |
| 2. Source context (Measure= 0 or 1) | - Does the source have a clearly stated aim/objective?<br>- Is the focus of study on RE in a large scale globally distributed agile setting?<br>- Are any limits clearly stated? |

| | | |
|---|---|---|
| | | - Is the source supported by documented references?<br>- Are the conclusions justified by the result? |
| 3. Publication Date<br>(Measure= 0 or 1) | | - Does the source have a clearly stated date? |
| 4. Significance of work<br>(Measure= 0 or 1) | | - Does the source enrich the current research, and/or particularly add something unique? |
| 5. Publication/literature type<br>(Measure= 0 to 4) | | - Academic peer reviewed: Extremely high credibility (Measure=4)<br>- PhD. /Master thesis: Very high credibility (Measure= 3)<br>- Peer reviewed experience reports: High credibility (Measure=2)<br>- Established vendor/leader (method creator) – case studies, blogs: Moderate credibility  (Measure=1)<br>- Commentary/opinion, blog posts from non-established vendor: Low credibility (Measure=0) |

Table 4.2: Quality criteria for study selection adopted from (Garousi et al., 2019)

## 4.7 Data extraction process

By following the guidelines of Cruzes and Dybå (2011), a data extraction process is defined to extract relevant information from the selected WL studies and GL studies. A standard data extraction form is created in an Excel spreadsheet, as shown in Appendix B. The following information will be extracted from each selected primary study.

Information of publication: Paper ID,

        Authors,

        Year of Publication,

        Title,

        Venue (where the source was published),

|                      | Quality score.                                                                 |
|----------------------|--------------------------------------------------------------------------------|
| Context descriptions: | Study Type- according to Wieringa et al. (2006) classification[2] (i.e. evaluative study, validation study, solution proposal, philosophical papers, opinion papers, experience papers), |
|                      | Settings (country/location of the analysis).                                   |
| Findings:            | Relevance to the theme, i.e., RE challenges, and RE strategies                 |

## 4.8 Data synthesis

Thematic synthesis will be used as a technique to synthesize the data as this technique organizes the material in such a way that it makes it easier to identify the key findings of the primary studies (Cruzes & Dybå, 2011). The thematic synthesis technique that will be used for this study will be inductive and data-driven, i.e., themes will have been proposed purely from the data extracted from the selected primary studies. The following steps will be undertaken to synthesize the data:

Step 1: Extract data

Step 2: Code data

Step 3: Translate codes into themes

Step 4: Reviewing themes

# 5. Validation of review process

How we validate our review process is described in this section.

Feedback on review protocol: First draft of the review will be circulated to Tony Clear and Ramesh Lal as research supervisors. The review protocol will be amended according to their feedback.

Pilot study: A pilot study will be conducted before conducting the main study to identify any problems in the review process, the review process will be amended accordingly.

---
[2] Detailed explanation of Wieringa et al. (2006) classification is provided in Appendix C.

Data extraction: A set of selected studies that would have been previously reviewed by the primary researcher will also be reviewed by the secondary and third researcher to cross check the extracted data.

Data synthesis: In order to validate the data synthesis, extracted themes will be cross checked with the secondary and third researcher as they are well experienced in the field of empirical software engineering, including secondary studies.

Major amendments to the protocol will be made in accordance with all feedback and reviews. The revised version will underpin the review. Should any further changes be required we will update this protocol and change the version number accordingly.

## 6. Reporting the review

It is planned to publish the process and results of performing the MLR on "*Requirements engineering (RE) in a global scaled agile software development environment*", in the journal *Information and Software Technology*. This will be supported by this detailed technical report to be published in the arXiv repository, *[https://arxiv.org/ - who assert that: arXiv is an open access resource. We are dedicated to the permanent custodial preservation of the scholarly record and to the rapid dissemination of scholarly scientific research]* that provides all the necessary transparency into the process and final reports.

## 7. Making changes to the Protocol

It is likely that changes to the protocol will be made when applying the procedures in new situations. Some changes will be made out of necessity, whereas other changes may be made to improve the current process. Every change to the protocol will be recorded and the protocol updated accordingly.

## Acknowledgements

This work has been undertaken in the course of the author's Doctoral Studies, with Associate Professor Tony Clear and Dr. Ramesh Lal as supervisors.  Their support for this work is acknowledged.

# Appendix A: Trial Search Record

Procedure to select final search string: First, we have developed three different search strings named search string 1, search string 2, search string 3 to retrieve the primary studies related to the phenomena of interest. Secondly, we performed the search on digital databases. Third, we compared the result of all search strings that we applied on digital databases. As a result of this, we finalised the search string 2 to retrieve the primary studies for this MLR study. The main reason to use the search string 2 as a final search string for this MLR study is that, it showed more consistent result as compared to search string 1 and search string 3.

**ScienceDirect**

**Search string 1[3]: articles retrieved 74**

( "requirements engineering" OR "user story" OR "business requirements" OR "high-level requirements" OR "requirements analysis" OR "development portfolio" OR "backlog management" ) AND ("distributed software development") AND ("LeSS")

**Search string 2: articles retrieved 62**

("requirements engineering" OR "requirements development" OR "requirements prioritization" OR "user story" OR "portfolio management" OR "backlog management" ) AND ("distributed software development") AND ("scaling agile" OR "LeSS")

**Search string 3 (which is a reverse string of search string 1): articles retrieved 62**

("scaling agile" OR "LeSS") AND ("distributed software development") AND ("requirements engineering" OR "requirements development" OR "requirements prioritization" OR "user story" OR "portfolio management" OR "backlog management" )

**SpringerLink**

**Search string 1: articles retrieved 0**

( "requirements engineering" OR "user story" OR "business requirements" OR "high-level requirements" OR "requirements analysis" OR "development portfolio" OR "portfolio management" OR "backlog management" ) AND ( global software* OR "distributed development" ) AND ( "scaled agile" OR "large-scale scrum" OR "disciplined agile delivery" )

---

[3] search string 1 that is mentioned above is same for all digital databases (e.g. search string 1 of ScienceDirect and search string 1 of IEEE are same). However, in each database, you will notice slight variation in search string 1 due to their acceptability of Boolean operators (e.g., ScienceDirect only accepts 8 Boolean operators whereas IEEE supports 10 Boolean operator). Therefore, search string 1 of ScienceDirect is slightly different than the search string 1 of IEEE.) Same procedure applied for search string 2 and search string 3 as well.

**Search string 2: articles retrieved 106**

( "requirements engineering" OR "requirements development" OR "requirements prioritization" OR "user story" OR "features" OR "portfolio management" OR "backlog management" ) AND ( "distributed software development" OR global software* ) AND ( "large-scale scrum" OR "scaling agile" OR "disciplined agile delivery" OR "scrum-of-scrum" )

**Search string 3 (which is a reverse string of search string 1): articles retrieved 97**

( "large-scale scrum" OR "scaling agile" OR "disciplined agile delivery" OR "scrum-of-scrum" ) AND ( "distributed software development" OR global software* ) AND ( "requirements engineering" OR "requirements development" OR "requirements prioritization" OR "user story" OR "features" OR "portfolio management" OR "backlog management" )

**Scopus**

**Search string 1: articles retrieved 46**

( "requirements engineering" OR "user story" OR "business requirements" OR "high-level requirements" OR "requirements analysis" OR "development portfolio" OR "portfolio management" OR "backlog management" ) AND ( global AND software* OR "distributed development" ) AND ( "scaled agile" OR "large scale scrum" OR "disciplined agile delivery" )

**Search string 2: articles retrieved 101**

( "requirements engineering" OR "requirements development" OR "requirements prioritization" OR "user story" OR "features" OR "portfolio management" OR "backlog management" ) AND ( "distributed development" OR global AND software* ) AND ( "large-scale scrum" OR "scaling agile" OR "disciplined agile delivery" )

**Search string 3 (which is a reverse string of search string 1): articles retrieved 101**

( "large-scale scrum" OR "scaling agile" OR "disciplined agile delivery" ) AND ( "distributed development" OR global AND software* ) AND ( "requirements engineering" OR "requirements development" OR "requirements prioritization" OR "user story" OR "features" OR "portfolio management" OR "backlog management" )

## ACM

**Search string 1: articles retrieved 13**

( "requirements engineering"  OR  "user story"  OR  "business requirements"  OR  "high-level requirements"  OR  "requirements analysis"  OR  "development portfolio"  OR  "portfolio management"  OR  "backlog management" )+ ( global  AND  software*  OR  "distributed development" ) +( "scaled agile"  OR  "large scale scrum"  OR  "disciplined agile delivery" )

**Search string 2: articles retrieved 36**

("requirements engineering" OR "requirements development" OR "requirements prioritization" OR "requirements management" OR "features" OR "backlog management" OR "user story" OR "portfolio management")+ ("global software development" OR "distributed software development" ) +( "large-scale agile" OR "large-scale scrum" OR "disciplined agile delivery ")

**Search string 3 (which is a reverse string of search string 1): articles retrieved more than 200000**

( "large-scale agile" OR "large-scale scrum" OR "disciplined agile delivery ")+ ("global software development" OR "distributed software development" ) +("requirements engineering" OR "requirements development" OR "requirements prioritization" OR "requirements management" OR "features" OR "backlog management" OR "user story" OR "portfolio management")

## IEEE

**Search string 1: articles retrieved 9**

((((((((((("requirements engineering")  OR  "user story")  OR  "business requirements")  OR  "high-level requirements")  OR  "requirements analysis")  OR  "development portfolio")  OR  "backlog management")  AND  "distributed software development")  AND  "scaled agile")  OR  "large scale scrum")  OR  "disciplined agile delivery"))

**Search string 2: articles retrieved 33**

((((((((((("requirements engineering") OR "requirements development") OR "requirements prioritization") OR "user story") OR "features") OR "portfolio management") OR "backlog management") AND "distributed software development) OR global software*) AND "large-scale scrum") OR "scaling agile"))

**Search string 3 (which is a reverse string of search string 1): articles retrieved more than 200000**

((((((((((("large-scale scrum") OR "scaling agile") AND "distributed software development) OR global software*) AND "requirements engineering") OR "requirements development") OR "requirements prioritization") OR "user story") OR "features") OR "portfolio management") OR "backlog management"))

# Appendix B: Data Extraction Form

| (Phase 1) | Researcher's Response | Comments |
|---|---|---|
| Paper ID | | ID of paper |
| Authors | | Name of authors |
| Year of publication | | Year of publication |
| Title | | Title of paper |
| Where paper was found (IEEE/ACM/RE proceedings etc.) | | Name of publishing authority |
| Date researcher analysed this paper | | When researcher completed this form |
| **Inclusion/Exclusion criteria** | | |
| Inclusion Criteria (a): Research Question/s answered? | | which research question/s is addressed by the paper |
| Inclusion Criteria (b): Acceptable source? | | We are not including Short paper/ panel discussions/ PowerPoint presentation/ poster in our study |
| Exclusion Criteria (a): Does study explicitly discuss requirements engineering and global scaled agile software development? | | |
| Exclusion Criteria (b): Is this a repeated study? | | We want to include only key study because repeated study will bias our result |
| Exclusion Criteria (c): Is this a SLR or tertiary study? | | SLR or Tertiary study will bias our result |
| **Quality Score** | | This will add trustworthiness in our study |
| **Decision** | | |
| Decision status {Accept/Reject/waiting for full paper} | | Define decision status |
| Decision based on: {Title/Abstract/Introduction/Conclusion/Method/Results/Whole Paper} | | At what point researcher take decision |
| **Context of study** | | |
| Study Type: Wieringa et al. (2006) classification | | Indicate type: evaluative research, validation research, solution proposal, philosophical papers, opinion papers, experience papers |
| For evaluative/empirical studies add: | | |
| Type of Empirical Study: {Questionnaire/survey; Interviews; Observation; Action research; Focus Groups}; | | |
| Country/Location of the analysis | | List countries involved in the study |
| **If a paper has passed all criteria in Phase1 above, Phase 2 (Qualitative Data Extraction) will be completed** | | |
| **Scaled agile requirements engineering issues and challenges** | | |
| **Challenge** | | |
| Challenge 1 (RQ1) | | |
| Challenge 2 (RQ1) | | |
| **Practices/strategies to overcome the challenge (models/methods/techniques)** | | |
| strategy for challenge (RQ2) | | |
| strategy for challenge (RQ2) | | |
| **Additional Data/Follow Up** | | |
| Other observations or useful quotes found in paper | | |
| References found in paper/snowballing (to follow up) | | |

# Appendix C: Study Type (according to Wieringa et al., 2006)

## Applied to study type field (In Data Extraction Form, Appendix B).

In order to determine the research type of the selected primary studies, we have adopted the Wieringa et al. (2006) classification scheme: evaluative research, validation research, solution proposal, philosophical papers, opinion papers, experience papers. A short description of each category is given in Table C1.

| Research type | Description |
|---|---|
| Evaluative research | This research presents the implementation and evaluation of a solution or technique in a real-world context, and the consequences are also investigated. Examples of these studies are case study, field study, survey, mathematics, etc. |
| Validation research | This research investigates the solution proposal that has not been implemented in practice. Possible research methods include simulation, laboratory experiments, etc. |
| Solution proposal | These papers propose the solution and discussed the potential benefits without full validation. |
| Philosophical papers | These papers sketch a new way of looking at things (e.g., a taxonomy or conceptual framework). |
| Opinion papers | Papers that reflect the personal opinion on a particular matter without relying on research methodologies and related work. |
| Experience papers | Papers that contain author's personal experience on what and how something has happened in practice. |

Table C1: Research type adapted from (Wieringa et al., 2006)